# EoSAlign: An Open–Source Software for Calculating, Comparing, and Aligning Pressure Under Extreme Conditions

Allison Pease[1*], Heidi N. Krauss[2], Kassandra Amezcua[1], Sang–Heon Shim[1]

*[1]School of Earth and Space Exploration, Arizona State University, Tempe, Arizona, United States of America; [2]Department of Earth and Environmental Sciences, Michigan State University, East Lansing, Michigan, United States of America*

*apease13@asu.edu

# EoSAlign: An Open–Source Software for Calculating, Comparing, and Aligning Pressure Under Extreme Conditions

## Abstract

Accurate pressure determination is essential in high pressure experiments. Inconsistencies among pressure scales limit direct comparisons and contribute to uncertainty in measured properties, such as, phase boundaries, melting conditions, and thermodynamic models, especially above 100 GPa where existing scales diverge by >10 GPa. EoSAlign, an open–source software, can calculate, compare, and align pressures using a range of published calibrations. EoSAlign calculates pressure from experimental measurements (such as volume (Å$^3$), Raman shift (cm$^{-1}$), or luminescence peak position (nm)) and it can be used to compare pressure values from different pressure standards. Using EoSAlign, we fit and align the bridgmanite and post–perovskite equations of state to a common scale, enabling us to reduce pressure scale related differences. EoSAlign can be applied to any high pressure data analysis to compare experimental results, improve consistency between studies, and advance our understanding of planetary interiors and high pressure physics.

## 1. Introduction

In high pressure experiments, pressure is often determined using an established pressure scale from an equation of state (EoS). However, not all published scales yield consistent pressure values, leading to discrepancies between studies at the 'same' conditions. Differences between studies arise from variations in a material's sensitivity to nonhydrostatic stress, equation-of-state formulation, fitted parameters, the reference standards used to define the pressure scale, and the degree of extrapolation beyond measured conditions. Even for the same pressure calibrant material, different research groups have reported pressure scales that diverge significantly at high pressures. In addition, some pressure scales in the same publications can differ at high pressure when more than one formulation or set of parameters is proposed for the same dataset. The use of different pressure scales creates a bottleneck when comparing studies: researchers must often reprocess or convert published pressure values into a common pressure scale for direct comparisons. This conversion process, which varies by group, is time–consuming and can introduce additional uncertainties or errors. As the number of high pressure data continues to grow, there is an increasing need for an extensive pressure scale database that makes it possible to easily compare and convert pressure scales.

EoSAlign, short for Equation of State Alignment, is an open–source software for calculating and comparing pressure at high pressures. EoSAlign serves as both a calculation software and a database of published pressure scales. EoSAlign currently contains published calibrations for Ag, Al, $Al_2O_3$, Ar, Au, Be, Bi, Co, Cu, diamond, Fe, He, KCl, MgO, Mo, NaCl, Ne, Ni, Pt, Re, $SrB_4O_7$,

Ta, W, YAG, and Zn; derived from X–ray diffraction, Raman spectroscopy, and luminescence studies. Using EoSAlign pressure can be calculated and compared from different inputs, such as pressure, volume, Raman shifts, or luminescence peak position. Pressure can be calculated by inputting measured experimentally measured values for one or more pressure scales enabling rapid pressure comparison during an experiment. Pressure can be compared between literature studies by inputting pressure selecting the published pressure scale and recalculating pressure based on the user input values for any published pressure scale. Within EoSAlign is EoSHolo, short for a holistic perspective of the EoS history. EoSHolo is a visualization tool that displays the interconnected network for each pressure scale. EoSAlign is designed to be continually updated with new pressure scales and enables users to upload their own calibration files into the software.

In this paper, we refer to multiple different published EoSs that differ due to the material and structure (e.g., MgO, Au, Fe–bcc, Fe–hcp), different parameters (e.g., K0 fixed vs. refined), different functional forms (e.g., Birch–Murnaghan vs. Rydberg–Vinet), different calibration histories (e.g., Ye et al., (1), Au intercalibrated to MgO vs. Pt), different pressure–transmitting media (PTM) (e.g., hydrostatic vs. non–hydrostatic), and different calibrations or combined datasets published by different research groups (e.g., (2,3)). We use the term pressure scale as a blanket label covering each of these differences. When referring to a specific difference between pressure scales, we apply the following notation: EoS composition for differences in material-structure-phase thus it does not refer to the thermodynamic definition, EoS parameters for differences in fitted parameter values, EoS formulation for differences in the functional equation, EoS history for differences in calibration history, EoS PTM for differences in the PTM, and published studies for differences between authors and datasets. We use the term "method" to refer to formulations linked to X–ray diffraction (XRD), luminescence and Raman spectroscopy. We also use the term calibration file to refer to how each published pressure scale is stored in EoSAlign.

## 2. Isothermal Equations of State and Pressure Proxies in EoSAlign

Here we review isothermal EoSs derived from static X–ray diffraction measurements and reduced shock isotherms, along with pressure proxies derived from spectroscopic observables measured using Raman spectroscopy and luminescence.

### *2.1 P0 term*

P0 represents the reference pressure. While it is typically ambient pressure, for phases which are not quenchable to 1 bar, some studies use the lowest stable pressures for reference pressure. In EoSAlign all, EoS, functions have a P0 term added. The reference pressure is assumed to be 0 unless specified by the published study in the calibration file.

### *2.2 Isothermal equation of state formulations*

The Murnaghan equation (4) assumed a linear pressure dependence of the bulk modulus; $K(P) = K_0 + K_0'P$.

$$P(V) = \frac{K_0}{K_0'}\left(\left(\frac{V_0}{V}\right)^{K_0'} - 1\right)$$

where $K_0$ is the isothermal bulk modulus, $K_0'$ is the first pressure derivative of the bulk modulus, $V_0$ is the volume. Subscript 0 indicates a reference condition.

The second–order Murnaghan EoS can take many forms, the form used in EoSAlign was reported in (5) to describe the pressure–volume relationship of Ne and Ar below 20 GPa.

$$P(V) = 2K_0 \frac{\left(\frac{V_0}{V}\right)^q - 1}{q\left(\left(\frac{V_0}{V}\right)^q + 1\right) - K_0'\left(\left(\frac{V_0}{V}\right)^q - 1\right)}$$

where $K_0$ is the isothermal bulk modulus, $K_0'$ is the first pressure derivative of the bulk modulus, $V_0$ is the volume at reference conditions, and $q = \sqrt{K_0'^2 - 2K_0K_0'')}$ (5).

The Birch–Murnaghan EoS (Birch, 1952, 1947) is derived from an Eulerian finite–strain expansion of Helmholtz free energy. The term "Murnaghan" in Birch–Murnaghan credits early nonlinear elasticity (Murnaghan, 1937) and is not a derivation from the Murnaghan EoS.

The third–order Birch–Murnaghan equation is:

$$\mathrm{P(V)} = \frac{3}{2}K_0\left(\left(\frac{V_0}{V}\right)^{\frac{7}{3}} - \left(\frac{V_0}{V}\right)^{\frac{5}{3}}\right)\left(1 + \frac{3}{4}(K_0' - 4)\left(\left(\frac{V_0}{V}\right)^{\frac{2}{3}} - 1\right)\right)$$

where $K_0$ is the isothermal bulk modulus, $K_0'$ is first pressure derivative of the bulk modulus, $V_0$ is the volume at reference conditions, and V is the observed volume at a given pressure. The second–order Birch–Murnaghan can be defined by setting $K_0' = 4$ and the fourth–order Birch–Murnaghan (e.g. (6)) can be defined by an additional expansion:

$$\mathrm{P(V)} = \frac{3}{2}K_0\left(\left(\frac{V_0}{V}\right)^{\frac{7}{3}} - \left(\frac{V_0}{V}\right)^{\frac{5}{3}}\right)\left(1 + \frac{3}{4}(K_0' - 4)\left(\left(\frac{V_0}{V}\right)^{\frac{2}{3}} - 1\right) + \frac{3}{8}\left(K_0K_0'' + (K_0' - 4)(K_0' - 3) + \frac{35}{9}\right)\left(\left(\frac{V_0}{V}\right)^{\frac{2}{3}} - 1\right)^2\right)$$

where $K_0''$ is the second pressure derivative of the bulk modulus at reference conditions.

The Rydberg–Vinet EoS (7) is derived from the universal binding–energy relation for solids (8) based on the interatomic potential from Rydberg (9). It is also called the 'Universal' EoS (e.g. (7,10)) and has recently been used more often in the geophysics community over the Birch–Murnaghan EoS to describe the extreme high pressure behavior of materials (e.g. (11)).

$$\mathrm{P(V)} = 3K_0 \frac{1-\left(\frac{V}{V_0}\right)^{\frac{1}{3}}}{\left(\frac{V}{V_0}\right)^{\frac{2}{3}}} \mathrm{e}^{1.5(K_0'-1)\left(1-\left(\frac{V}{V_0}\right)^{\frac{1}{3}}\right)}$$

An alternative formulation is the extended (higher–order) Rydberg–Vinet EoS, this is commonly used to describe compression at very high pressures (e.g. (12–15)).

$$P(V) = 3K_0 \frac{1-\left(\frac{V}{V_0}\right)^{\frac{1}{3}}}{\left(\frac{V}{V_0}\right)^{\frac{2}{3}}} \mathrm{e}^{\eta\left(1-\left(\frac{V}{V_0}\right)^{\frac{1}{3}}\right)+\beta\left(1-\left(\frac{V}{V_0}\right)^{\frac{1}{3}}\right)^2+\psi\left(1-\left(\frac{V}{V_0}\right)^{\frac{1}{3}}\right)^3+\gamma\left(1-\left(\frac{V}{V_0}\right)^{\frac{1}{3}}\right)^4}$$

where η is the leading–order exponential coefficient analogous to the pressure derivative of the bulk modulus ($K_0'$) in the standard Rydberg–Vinet equation, and β, $\psi$, and γ are higher–order coefficients that account for deviations from the standard Vinet form at high pressure.

The adapted polynomial (AP) and the Holzapfel EoS extend the Rydberg–Vinet formulation by enforcing the asymptotic Thomas–Fermi limit under extreme compression. These formulations were organized into two series, H0L and H1L, where L denotes the polynomial order. The H0L series places polynomial correction terms inside the exponential but does not include a Fermi–gas parameter and diverges at high pressure; the H1L series applies the Fermi–gas parameter, with the polynomial correction appearing inside the exponential. The second–order members of these series are H02 and H12, respectively. H02 is the form used (16,17) in their Ne, Ar, Al, and Au pressure scales. In the early 2000s, Holzapfel reformulated H12 by moving the polynomial correction outside the exponential; this version is AP2, and it is the form used in (18–20).

H02:

$$\mathrm{P(V)} = 3K_0 \frac{1-\left(\frac{V}{V_0}\right)^{\frac{1}{3}}}{\left(\frac{V}{V_0}\right)^{\frac{5}{3}}} \mathrm{e}^{\left(\frac{3}{2}(K_0'-3)\left(1-\left(\frac{V}{V_0}\right)^{\frac{1}{3}}\right)\right)}$$

H12:

$$\mathrm{P(V)} = 3K_0 \frac{1-\left(\frac{V}{V_0}\right)^{\frac{1}{3}}}{\left(\frac{V}{V_0}\right)^{\frac{5}{3}}} \mathrm{e}^{\left(-ln\left(\frac{3K_0}{2337}\right)-\frac{5}{3}\ln\left(\frac{V_0}{Z}\right)\right)\left(1-\left(\frac{V}{V_0}\right)^{\frac{1}{3}}\right)+\left(\frac{3}{2}(K_0'-3)+\left(ln\left(\frac{3K_0}{2337}\right)+\frac{5}{3}\ln\left(\frac{V_0}{Z}\right)\right)\right)\left(\frac{V}{V_0}\right)^{\frac{1}{3}}\left(1-\left(\frac{V}{V_0}\right)^{\frac{1}{3}}\right)}$$

AP2:

$$\mathrm{P(V)} = 3K_0 \frac{1-\left(\frac{V}{V_0}\right)^{\frac{1}{3}}}{\left(\frac{V}{V_0}\right)^{\frac{5}{3}}}\left(1+\left(\frac{3}{2}(K_0'-3)+\left(ln\left(\frac{3K_0}{2337}\right)+\frac{5}{3}\ln\left(\frac{V_0}{Z}\right)\right)\right)\left(\frac{V}{V_0}\right)^{\frac{1}{3}}\left(1-\left(\frac{V}{V_0}\right)^{\frac{1}{3}}\right)\right)\mathrm{e}^{\left(-ln\left(\frac{3K_0}{2337}\right)-\frac{5}{3}\ln\left(\frac{V_0}{Z}\right)\right)\left(1-\left(\frac{V}{V_0}\right)^{\frac{1}{3}}\right)}$$

where 2337 GPa $Å^5$ is the Fermi–gas parameter, and Z is the atomic number (20).

The Keane EoS (21) requires that the pressure derivative of the bulk modulus, K′, decreases with pressure toward a finite limit $K'_{\infty}$ at infinite pressure.

$$P(V) = K_0\left(\left(\frac{K_0'}{{K_{\infty}'}^{2}}\right)\left(\left(\frac{V_0}{V}\right)^{K_{\infty}'} - 1\right) - \left(\frac{K_0'}{K_{\infty}'} - 1\right)\ln\left(\frac{V_0}{V}\right)\right)$$

where $K'_{\infty}$ is the pressure derivative of the bulk modulus at infinite pressure.

In 2016, Sakai et al. (22) applied both the AP2 and Keane EoS, to describe material behavior at extreme compressions where the Rydberg–Vinet and Birch–Murnaghan equations reported for the same experimental dataset began to diverge. Recently, Sakai et al., (23) applied the Rydberg–Stacey EoS (24). This EoS is similar to the Rydberg–Vinet EoS. Where the pressure derivative at infinite pressure can be fit using values of 2/3, 5/3, and 3 result in behavior similar to the Rydberg–Vinet, AP2, and Birch–Murnaghan ($3^{rd}$ order), respectively.

$$\mathrm{P(V)} = 3K_0\left(\frac{V}{V_0}\right)^{-K'_{\infty}}\left(1 - \left(\frac{V}{V_0}\right)^{\frac{1}{3}}\right)\mathrm{e}^{\left(\frac{3}{2}K'_0 - 3K'_{\infty} + \frac{1}{2}\right)\left(1 - \left(\frac{V}{V_0}\right)^{\frac{1}{3}}\right)}$$

### *2.3 Pressure dependent changes in luminescence peak position*

This group of pressure scales includes ruby fluorescence and the luminescence of other common materials such as YAG and $SrB_4O_7$. Five formulations have been proposed to explain the shift in wavelength as a function of pressure.

The first relationship is a linear scale:

$$P(\lambda) = \frac{\lambda - \lambda_0}{A}$$

where $\lambda_0$ is ambient peak position and A is the slope representing how the peak position changes with pressure. The linear scale was historically used in initial ruby fluorescence studies (25,26) and is commonly used for materials (e.g. spinel, tourmaline, zircon, alexandrite) that do not require higher order expansions, typically valid at low pressures (e.g. (27)).

A common expression for the ruby shift, used in (28–32) among others was based on a modification to the Murnaghan EoS form.

$$P(\lambda) = \frac{A}{B}\left(\left(\frac{\lambda}{\lambda_0}\right)^{B} - 1\right)$$

where A and B are fitting parameters, $\lambda_0$ is the ambient position, and $\lambda$ is the position at a given pressure.

The Murnaghan form has been demonstrated to be a good fit to observations at moderate pressures, but alternative expressions have been derived to explain the high pressure behavior of ruby. A relationship proposed in Aleksandrov et al., (33) was recently used in (1,34,35). This equation has been proposed to better agree with experimental and theoretical observations (33) at high pressure.

$$P(\lambda) = A\,\frac{\lambda - \lambda_0}{\lambda_0}\left(1 + B\frac{\lambda - \lambda_0}{\lambda_0}\right)$$

where A and B are fitting parameters, $\lambda_0$ is the ambient position, and $\lambda$ is the position at a given pressure.

A relationship proposed by Holzapfel (18,19,36) uses the third–order modified Freud–Ingalls form. This expression matches the slope and curvature of the Murnaghan form at low pressure but adopts a different curvature at high pressure (parameter C).

$$\lambda(P) \;=\; \lambda_0\left(1 - \left(\frac{C}{B+C}\right)\ln\left(1 + \left(\frac{B+C}{A}\right)P\right)\right)^{-\frac{1}{C}}$$

$$P(\lambda) = \frac{A}{B+C}\left(e^{\left(\frac{B+C}{C}\left(1 - \frac{\lambda_0}{\lambda}^{C}\right)\right)} - 1\right)$$

where A, B, C are fitting parameters, $\lambda_0$ is the ambient position, and $\lambda$ is the position at a given pressure.

Datchi et al., (37) formulated a different expression for $SrB_4O_7$.

$$P(\lambda) = A\Delta\lambda\frac{1 + B\Delta\lambda}{1 + C\Delta\lambda}$$

where A, B, C are fitting parameters and $\Delta\lambda = \lambda - \lambda_0$.

### *2.4 Pressure dependent changes in Raman shift*

This group of pressure scales includes the splitting of the diamond edge and the Raman shift of materials (Be, Re, cBN). Numerous expressions have been used to explain the Raman shift of the materials and at least four have been applied to the splitting of the diamond edge.

To calculate pressure from the splitting of the diamond edge, Akahama and Kawamura (38) proposed a polynomial formulation

$$P(\upsilon) = A + B\upsilon + C\upsilon^2$$
$$\upsilon(P) = D + E(P + F)^{\frac{1}{2}}$$

where $\upsilon$ is the position in relative wavenumbers, P is pressure, and A–F are constants fit to the data.

The second equation was used in Akahama and Kawamura (39), this was based on the finite strain approximation and was also used by Aleksandrov et al., (33).

$$P(v) = K_0\left(\frac{v - v_0}{v_0}\right)\left(1 + \frac{(K_0' - 1)\left(\frac{v - v_0}{v_0}\right)}{2}\right)$$

The third equation was proposed by Akahama and Kawamura (40). In this study they observed a divergence above 300 GPa between the Akahama and Kawamura (39) pressure scale and their new high pressure data. Therefore, they applied the equation presented in Akahama and Kawamura (39) below 200 GPa and add a quadratic expression for values conditions > 200 GPa.

$$P(v, if\ < 200\ GPa) = A\left(\frac{v - v_0}{v_0}\right)\left(1 + \frac{(B - 1)\left(\frac{v - v_0}{v_0}\right)}{2}\right)$$

$$P(v, if \geq 200\ GPa) = C - Dv + Ev^2$$

where $v$ is the position in relative wavenumbers, $v_0$ is the ambient position, P is pressure, and A–F are constants fit to the data.

The fourth equation was proposed by Eremets et al., (41) and was later used by Sakai et al., (23). This formulation provides a good fit from ambient to > 400 GPa and is the same formulation used in Akahama and Kawamura (39) where A and B are defined differently.

$$P(v) = A\left(\frac{v - v_0}{v_0}\right) + B\left(\frac{v - v_0}{v_0}\right)^2$$

The remaining formulations are specific to the Raman shift of Be, Re, or cBN.

For Be, Evans et al., (42) applied the following expression:

$$v(P) = A + BP + CP^2$$

For Be and Re, Olijnyk et al., (43) applied the following expressions:

$$v(P) = v_0\left(1 - \frac{B}{A}P\right)^{-\frac{A^2}{B}}$$

$$P(v) = \frac{A}{B}\left(1 - \left(\frac{v}{v_0}\right)^{-\frac{B}{A^2}}\right)$$

For Re, Pease et al., (44) applied the following expression:

$$P(v) = A(v - v_0) + B(v - v_0)^2$$

For cBN, Goncharov et al., (45,46) applied the following expression:

$$P(v) = \frac{A}{B}\left(\left(\frac{v}{v_0}\right)^B - 1\right)$$

While, others (47–49) applied one/both of the following polynomial expressions depending on if temperature was considered:

$$v(P,T) = v_0 + AT + BT^2 + (C + DT + ET^2)P + FP^2$$

$$v(P) = v_0 + AP + BP^2$$

And Datchi et al.,(50) applied the following expression:

$$P(\upsilon) = \frac{A}{3.62}\left(\left(\frac{\upsilon}{\upsilon_0}\right)^{2.876} - 1\right)$$

Where A=$K_0$; $K_0' = 3.62, \gamma = 1.257$

$$\upsilon(P) = \upsilon_0\left(1 + \left(\frac{K_0'}{K_0}\right)P\right)^{\frac{\gamma}{K_0'}}$$

In each expression for Be, Re, cBN, $\upsilon$ is the peak position in relative wavenumbers, $\upsilon_0$ is the ambient position, P is pressure, and A–F are constants fit to the data, T is the temperature dependence, in EoSAlign this is assumed to be room temperature, 300 K. In some studies (e.g., (47,50)) the fitted constants are directly related to the P–V–T EoS. Specifically, in Datchi et al., (50) parameter A is the isothermal bulk modulus, 3.62 is the first pressure derivative of the bulk modulus, and 2.876 is $\frac{K_0'}{\gamma}$, linking the Raman modes to the materials elastic properties using a Murnaghan formulation.

## 3. Pressure scales in EoSAlign

The goal of EoSAlign is to serve as a database for a large number of published pressure scales so that anyone working with high pressure data can easily convert and compare experimental data. The EoSAlign database can be constantly updated by users or developers. It currently consists of more than 400 pressure scales. Each published pressure scale is saved as a calibration file and contains metadata about the scale. Users can edit these calibration files and add their own pressure scales. New pressure scales can be added through the menu bar by navigating to "File" and then "Enter Calibration". It is recommended to enter a calibration file using this method to avoid formatting issues. Table 1 summarizes the pressure calibrant materials and scales that can currently be found in the database.

[Table 1 near here]

To build the EoSAlign database, we extracted the EoS parameters from tables in the literature. It is possible that some manuscripts could contain typos in their reported EoS parameters or data points. The problem can be identified during refitting the data, because the typos could result in EoS parameters that differ from the reported values, leading to systematic errors in the literature. To ensure that these errors can be tracked and realigned in studies that cite these pressures scales we do not refit the published EoS parameters, when applicable, all pressure scales will appear in EoSAlign as they were published.

Unfortunately, some pressure scales do not report all fitting parameters. To document this problem, in EoSAlign the scales are grouped into one of three categories. Category 1, no alteration to the pressure scale was made, and they were originally reported in the publication. Category 2, we made a minor alteration. This includes cases where either $V_0$, $K_0$, or $K_0$' are not directly reported in the original publication and therefore needed to be calculated from provided information. This category also includes cases where the volume, Raman shifts, or luminescence peak position at the reference conditions was not defined in the original publication. In this case, the values are assumed based on another study cited within the calibration file. Category 3, a 3$^{rd}$ order Birch–Murnaghan EoS was fit to a published data table. Some studies, which are used to calibrate modern pressure scales, only report the raw data or reduced shock isotherms rather than an isothermal 300 K EoS to the data. Table 2 reports Birch–Murnaghan (3$^{rd}$ order) fits to the published data used in EoSAlign.

[Table 2 near here]

## 4. Workflow

The EoSAlign software can be accessed through the associated GitHub repository where users can choice between the Python code files, desktop applications, a pip install package, or a web version of the software. When the application initiates the user is prompted to select one of the available features which currently include EoSAlign and EoSHolo. All features are accessible at any time through the menu bar under the "Application" option.

### *4.1 EoSAlign*

EoSAlign is the primary feature. Its purpose is to calculate pressure from a given pressure scale and convert pressure between different pressure scales. The workflow and optional paths for EoSAlign are shown in Figure 1. EoSAlign is divided into two primary paths, pressure calculation and pressure alignment. The path can be chosen based on input property which can be entered manually or uploaded as a txt file. Once input parameters are entered, a unit can be selected, such as pressure (GPa), wavelength (nm), relative wavenumbers (cm$^{-1}$), or volume (Å$^3$/unit-cell, Å$^3$/formula unit, Å$^3$/atom, $cc^3$/mol) for the input parameters. Error propagation is optional. If error propagation is selected, a user needs to repeat the process above, ensuring units and number of data points match. Next a user selects the pressure calibrant composition. Once a pressure scale is selected, the path will split depending on if the properties entered at the start were pressure or volume/Raman shift/luminescence peak position.

PATH 1. If the starting units were volume/wavelength/relative wavenumbers EoSAlign is set up to calculate pressure based on the measured input properties. This path is designed for calculating pressure and comparing different pressure scales for given inputs. Next, a user selects the scale(s) to calculate pressure. It is recommended to click the 'Preview Calibrant' button to obtain more information on the selected pressure scale. The calibration file can be edited if necessary. Viewing the file can be used to examine if uncertainties are assigned to the parameters. If they are not, then uncertainty will be underestimated. If a user has measured input parameters at ambient pressure, the ambient pressure property from published EoSs can be overwritten temporarily using the adjust reference button.

PATH 2. If the input value was pressure, the software will compare the input pressure to those from other pressure scales. This is achieved by converting the input pressure to volume, luminescence peak wavelength, or Raman mode frequency. Then using the input, EoSAlign can recalculate and compare pressure from other calibrants. Therefore, the next step is to define the pressure scale that input data was calculated from. After this step, there are two paths available. The first path (2a) calculates pressure based on a study with the same composition and method. The second path (2b) calculates pressure based on a study with a different composition and method.

PATH 2a. The composition and method will match those of the original input. A user will need to select the pressure scales they would like the pressure to be converted to.

PATH 2b. The composition and method do not match ones used for the original input. A user will need to define the composition and method they wish to convert to; not all combinations of pressure calibrants are possible; this path requires a study that reports pressure scales for multiple pressure calibrants and assumes that all calibrants in the study were collected under the same conditions. This assumption of equal pressure of multiples of pressure calibrants may introduce uncertainty but is a first step toward pressure comparison between studies calibrated with different pressure calibrants. Therefore, only intercalibration studies that report a pressure scale of the input and the desired conversion pressure scale appear as conversion points; studies that calibrate their study to another scale are not included. There are limits to this, for example, Dorfman et al.,(75) intercalibrated numerous pressure calibrants to MgO using the MgO pressure scale by Tange et al.,(121) but Dorfman et al., (75) did not report a new pressure scale for MgO. Therefore, a direct conversion between Au and MgO through Dorfman et al.,(75) is not supported and will need to be done manually by the user. This restriction was implemented to prevent an excess of assumption paths that could develop in studies calibrated to multiple pressure scales (e.g. (35)) ensuring a more direct comparison.

At the end of each path three buttons will appear; 'Plot Results', 'Preview CSV Data', and 'Export Results'. The 'Preview CSV Data' button will display all input and converted values in a table. The 'Export Results' button will save the input and converted values as a CSV file. The 'Plot Results' button will display multiple plots of the input and converted values. The plots use solid symbols and lines when the values are within the pressure range of the corresponding calibration. Open symbols and dashed lines are used when the values are outside of the calibration's specified pressure range. If a calibration did not report a pressure range, they will always appear as open symbols and dashed lines.

[Figure 1 near here]

### *4.2 EoSHolo*

EoSHolo creates a visualization of the interconnected network of pressure calibrations. Each calibration is given a representative point, and arrows are used to visualize the connections between the calibrations. These representative points are initially placed within a column that represents the composition or structure used for the calibration. The user can move these points around to fully visualize the entire network. All calibrations included in this software are represented by grey circles while calibrations not currently included are represented by red triangles. At any point a user can select a specific calibration by either right clicking the corresponding point or selecting a composition and calibration. When a specific calibration is selected it will appear as a star and all other pressure calibrations that went into the creation of the selected calibration will appear as blue circles. Users can right click on any calibration to get more information, view the calibration information, and view the calibration's origin chain or view the complete chain. The calibration's origin chain button is the default when a calibration is selected. The calibration's complete chain button function displays the origin and all calibrations that have built off the selected calibration as green circles. The 'Show All Nodes' checkbox allows users to display all possible calibrations when checked or only display the nodes connected to the selected calibration when unchecked. This function allows users to visualize all direct and indirect calibration relationships providing a clear view of how pressure calibrations are all interconnected.

### *4.3 Citation*

When using EoSAlign to convert, calculate, or align pressure, we recommend that the conversion is trackable by citing each path in the following formats.

1. Pressure for (insert composition) was calculated using EoSAlign (Pease et al., 2026, version #) from (insert study name), (insert equation of state) with (identifying term if needed).
   *Example: Pressure for Au was calculated using EoSAlign (Pease et al., 2026, version#) from Dorfman et al., 2012,* Rydberg–*Vinet Equation of State, with K0 fixed.*

2. Pressure was recalculated for (insert composition) using EoSAlign (Pease et al., 2026, version #) from (insert study name), (insert equation of state) with (identifying term if needed) to (insert study name), (insert equation of state) with (identifying term if needed).
   *Pressure was recalculated for Au using EoSAlign (Pease et al., 2026, version #) from Dorfman et al., 2012,* Rydberg–*Vinet Equation of State, with K0 fixed to Sakai et al., 2025,* Rydberg–Stacey equation of state.

3. Pressure was recalculated for (insert composition 1) to (insert composition 2) using EoSAlign (Pease et al., 2026, version #) from (insert study name), (insert equation of state) with (identifying term if needed) through (insert study name), (insert equation of state) with (identifying term if needed) to (insert study name), (insert equation of state) with (identifying term if needed).
   Pressure was recalculated for Au to Pt using EoSAlign *(Pease et al., 2026, version#)* from Shim et al., 2002, Birch–Murnagham *Equation of State, with K0 fixed,* through Dorfman et al., 2012, Rydberg–*Vinet Equation of State, with K0 fixed,* to Ye et al., 2017, Rydberg–*Vinet Equation of State, calibrated from MgO.*

Studies typically publish multiple pressure scales with different EoS formulations, parameters, PTM, or pressure range. To ensure that future studies are directly comparable with the literature and minimize calibration correction uncertainty, it is recommended to ensure the pressure calibration used is trackable whether EoSAlign is used for the conversion or not.

## 5. Pressure Difference Between Select Pressure Scales

EoSAlign currently supports 32 different "compositions" (materials or phases) calibrated using three different methods. Some materials have dozens of reported pressure scales. EoSAlign can export calculated values and generated editable plots in standard formats. Plots open in a new window, the first tab enables studies to be toggled on and off, enables a user to change the marker symbol/color and preview the calibration file. The next few tabs contain the figures: measure values (e.g. volume, luminescence peak positions, Raman shift frequencies) vs pressure, selected studies $P_{diff}$ (pressure differences) vs pressure, selected studies $P_{diff}$ % vs pressure, absolute and relative pressure spread, one selected studies' $P_{diff}$ vs pressure, one selected studies' $P_{diff}$ % vs pressure, and a combination figures (as exported in SI material). Any permutation of figures can be generated for pressure scales within EoSAlign. Due to complexities and interactive nature of these plots, it is not feasible to add them directly into this paper. Therefore, combination figures generated for "all scales" have been added to the supplementary material when there are more than three pressure scales listed in EoSAlign.

[Figure 2 near here]

Gold is one of the most used materials for pressure measurements in DAC. More than 35 pressure calibration files are currently available on EoSAlign. Figure 2 summarizes the range in the published pressure scales of Au up to 500 GPa. Many of these are extrapolated to higher pressures than the pressure range they are calibrated to. The ~120 GPa spread in the data at 500 GPa with a standard deviation of ~40 GPa reflects all studies including extrapolated values. Recent pressure scales that report a max pressure around or greater than 500 GPa are in better agreement. Figure 3 illustrates the pressure difference with violin plots between the most used pressure scales with their distributions: Au, MgO, Pt, Re with X–ray diffraction, diamond with Raman, and ruby with luminescence. For each pressure scale the discrepancy between individual scales increases with pressure.

[Figure 3 near here]

For each of the commonly used pressure scales, the pressure difference increases with increasing pressure. When all studies in EoSAlign are considered (extrapolated and non–extrapolated), the spread at core–mantle boundary pressures (136 GPa) has an interquartile range (IQR, the spread of the middle 50% of values, from the 25th to the 75th percentile) of 4 to 9 GPa with outliers extending to ±15 GPa across scales. Median differences and IQRs are: Au (median 1.7 GPa, IQR 4.0 GPa), MgO (0.1 GPa, 8.9 GPa), Pt (1.5 GPa, 3.9 GPa), Re (3.8 GPa, 6.9 GPa), Diamond (7.6 GPa, 6.4 GPa), and Ruby (–0.1 GPa, 7.2 GPa). Diamond and Re show the largest systematic offsets from the reference, while MgO and Ruby exhibit the largest outliers, with MgO ranging from –11.9 GPa to +31.5 GPa and Ruby from –30.3 GPa to +6.1 GPa. At core pressures (360 GPa), the spread broadens dramatically. IQRs reach 10 to 42 GPa, and select studies overestimate the pressure by more than 100 GPa (Figure 3). Median differences and IQRs are: Au (–0.2 GPa, 20.2 GPa), MgO (–4.0 GPa, 42.0 GPa), Pt (2.3 GPa, 23.9 GPa), Re (17.8 GPa, 23.8 GPa), and Diamond (0.5 GPa, 10.3 GPa). Ruby is not reported at this pressure because fluorescence cannot be probed. MgO and Re have the largest outliers at +227 GPa and +145 GPa, respectively, and Re also shows the largest median offset.

The disagreement decreases when only non–extrapolated studies are considered. At 136 GPa, the in–range medians and IQRs are: Au (2.2 GPa, 4.6 GPa), MgO (0.3 GPa, 3.1 GPa), Pt (2.1 GPa, 3.0 GPa), Re (6.6 GPa, 7.3 GPa), Diamond (7.6 GPa, 3.8 GPa), and Ruby (0.0 GPa, 0.8 GPa). The IRQ of ruby decreased to <1 GPa, and MgO dropped from 8.9 to 3.1 GPa, indicating that extrapolation is the dominant source of discrepancy at core mantle boundary pressures for these scales. At 360 GPa, in–range data are sparse (2 to 11 studies per scale), but the discrepancy also decreases: Au (1.8 GPa, 8.8 GPa), MgO (–4.0 GPa, 2.6 GPa), Pt (20.4 GPa, 20.2 GPa), Re (20.3 GPa, 5.8 GPa), and Diamond (1.7 GPa, 22.8 GPa). Platinum and Re have large medians (~20 GPa), suggesting that discrepancy between scales at core pressures is not purely an artifact of extrapolation but reflects genuine differences in how the scales are calibrated and what the experimental conditions are.

Some high pressure studies conducted intercalibration for pressures calculated from multiples of materials (e.g. (2,23,29,75,82,86)). In these studies, pressure is calibrated from multiples of materials loaded together in a diamond anvil cell, multiple experiments on different EoS compositions by the same research group, or the merger of multiple datasets between research groups. The intercalibration studies make it possible to constrain the high pressure behavior of multiple pressure calibrant materials simultaneously as the materials are assumed to be at the same condition (pressure, temperature, differential stress). Assuming that all the materials in the sample are under the same conditions, these intercalibration studies are the best way to relate different pressure scales to one another, and they enable Path 2b in EoSAlign. However, intercalibration studies are not equivalent, each assumes a particular absolute pressure scale as the reference to which the other compositions are aligned. For demonstration, we compare six intercalibration studies and align their reported Au pressure scales to a recent Cu standard from Fratanduono et al., (13).

[Figure 4 near here]

Figure 4a shows the pressure difference between the studies and Sakai et al., (23) at 10, 50, and 136 GPa. Using the published Au pressure scales without conversion, the studies agree well at low pressure but diverge at core–mantle boundary pressure. At 10 GPa the median pressure difference between the nine studies is 0.45 GPa (IQR 0.03 GPa, range 0.30 to 0.67). At 50 GPa it is 0.83 GPa (IQR 0.38 GPa, range 0.24 to 1.55). At 136 GPa the median pressure difference is low (0.49 GPa), but the IQR widens to 1.78 GPa and individual pressure difference increases to –4.34 GPa to 3.91 GPa. Within this set, Shen and Smith (82) and Sakai et al., (23) agree within 1 GPa to 160 GPa. Ye et al.,(1) and Dorfman et al.,(75) agree to within ~2 GPa and the Fei et al., (2) and Dewaele et al.,(29) scales differ by roughly 4 and 6 GPa respectively, although these comparisons are made beyond their calibration range.

Of the intercalibrated Au pressure scales, the studies can be divided into three types. The first type is no conversion needed, for example, Sakai et al., (23) and Shen and Smith (82) because they both are calibrated to Fratanduono et al., (13). The second type is the study that was originally calibrated to a Cu pressure scale (in some capacity); this occurs for Fei et al., (2) and Dewaele et al.,(29). Fei et al., (2) references Dewaele et al., (29) and both cite additional multi–calibrant papers, but each links back to Mao et al., (30) ruby scale which was calibrated to Carter et al., (52), using both the Ag and Cu scale. Because separating the Ag vs. Cu contribution is not clear, we assume that the Mao et al., (30) ruby pressure yields the same pressure as the Carter et al., (52) Cu pressure scale, and then convert the pressure based on the Carter et al., (52) Cu scale to that based on the Fratanuono et al., (13) Cu scale. After this conversion, the scale by Fei et al., (2) shows improved agreement with Sakai et al., (23) at 160 GPa, while the Dewaele et al., (29) scale becomes increasingly negative, suggesting the mismatch may originate from individual datasets within those studies. The third type is conversion through a different composition; this occurs for the scales by Ye et al., (1) and Dorfman et al., (75). The Dorfman et al., (75) scale was intercalibrated with MgO to Tange et al., (121) and the Ye et al., (1) scale reported two intercalibrations to the MgO and Pt (34) scale. Neither study has a calibration link directly to a Cu pressure scale. Therefore, the conversion to the Fratanduono et al., (13) scale must propagate through a different pressure calibration study, such as Sakai et al., (23) and Shen and Smith (82) each of which reported pressure scales for Cu, MgO, and Pt.

At 136 GPa, the choice of conversion study produces different results, and in most cases the conversion increases the discrepancy with the Cu scale. For Ye et al., (1) when MgO intercalibrated, the original pressure difference among the scales studied there is +1.7 GPa. Routing through the scales in Sakai et el., (23) increases it to +6.5 GPa (~4.8 GPa worse), while routing through the scales in Shen and Smith (82) decreases it to ~0.0 GPa. This is the only case among the eight MgO/Pt–intercalibrated conversions where the conversion study significantly decreases the pressure difference through routing. For the Dorfman et al. ((75), MgO–intercalibrated) scales, the original offset is +0.5 to +1.3 GPa depending on the EoS parameters in the study. Converting through the Sakai et al., (23) scales increases the pressure difference to –1.4 to –2.2 GPa, while the Shen and Smith (82) conversion leads to a –7.2 to –8.0 GPa pressure difference, leading to the largest disagreement in the dataset. For Ye et al., ((1), Pt–intercalibrated), the original pressure difference is ~0.1 GPa this expands to –2.1 GPa when converting through Sakai et al., (23) and –0.3 GPa through Shen and Smith (82).

This pattern is consistent with Shen and Smith (82) and Sakai et al., (23) for the MgO and Pt intercalibrations in that they differ by a few GPa at 136 GPa: the MgO scale by Shen and Smith (82) is ~6 GPa lower than the MgO scale by Sakai et al.,(23), while the Pt scale by Shen and Smith (82) is ~2.5 GPa greater than that of Sakai et al., (23) at 136 GPa, even though their Cu scales align well within 200 GPa. The conversion study disagreement is likely linked to the pressure–transmitting medium and therefore the hydrostaticity of each experiment. Shen and Smith (82) is quasi–hydrostatic while Sakai et al., (23) is more non–hydrostatic. Studies may align better through conversion studies that most closely matches their own experimental conditions. Dorfman et al., (75) used Ne and NaCl, of which NaCl is closer to non–hydrostatic, and consequently their EoS aligns better through Sakai et al., (23) (–1.4 to –2.2 GPa) than through Shen and Smith (82) (–7.2 to –8.0 GPa). The opposite holds for Ye et al., ((1), which used Ne and Ar (closer to hydrostatic to quasi–hydrostatic) which aligns better through Shen and Smith (82) for both the MgO–intercalibrated version (+0.0 vs +6.5 GPa (23)) and the Pt–intercalibrated version (–0.3 vs –2.1 GPa (23)).

Surprisingly, the realignment of the Au pressure scales to a common Cu standard (13) increased the pressure difference relative to the unconverted pressure scale. At 10 GPa, the original median pressure difference is 0.45 GPa and the IQR is 0.03 GPa (range 0.30 to 0.67 GPa); after conversion, the IQR grows to 0.13 GPa and the range widens (–0.14 to 0.59 GPa). At 50 GPa, the original IQR is 0.38 GPa (range 0.24 to 1.55 GPa), and the converted IQR is 0.63 GPa (range –0.76 to 1.72 GPa). At 136 GPa, the original median is 0.49 GPa with an IQR of 1.78 GPa (range –4.34 to 3.91 GPa), while the converted median shifts to –1.38 GPa with an IQR of 4.59 GPa (range –7.97 to 6.50). The comparison between the original and converted scales reveals the true variation among the Au pressure scales and underscores the importance of experimental conditions (particularly hydrostaticity) when selecting a pressure scale.

## 6. Error Propagation

Error propagation is important in comparing results between studies. The error propagation in EoSAlign is calculated using a gaussian error propagation where the partial derivatives are calculated using the central difference method with a step of 0.0001. Error propagation is calculated once to determine the uncertainty in pressure ($\sigma_P$) when the user enters measured values in volume, Raman shift, or luminescence peak position (path 1). The error propagation is then calculated two (path 2a) to four (path 2b) times when the user enters pressure. Figure 5 is an example of error propagation from an Au pressure scale to W pressure scales. Notice how with the error propagation the three W pressure scales are at lower pressure than the Au scale (75) when converted through Sakai et al., (23). The error propagation calculates uncertainties to pressure comparisons and calculations. However, it is important to note that some studies do not report uncertainties for their EoS parameters; therefore, the error propagation function is set up as optional. It is recommended to only propagate the uncertainty for pressure scales where their uncertainty is reported and added to the calibration file because if the uncertainty is not provided it is assumed to be zero and will lead to an underestimation of pressure uncertainty.

[Figure 5 near here]

## 7. An Example Application of EoSAlign: Equations of State of Post–Perovskite and Bridgmanite

Over the past decades, researchers have investigated both endmember and realistic compositions of major mineral phases expected for Earth and planetary interiors. However, these studies often report different thermodynamic parameters, elastic parameters, and phase transition boundaries. While these differences can arise from the starting materials used (e.g., purity, synthesis conditions, and water content), and experimental methods (e.g., strain rate and deviatoric stress), the choice in pressure scale used for calibration also contributes significantly.

Bridgmanite is the most abundant mineral in the Earth's interior, and its high pressure polymorph, post–perovskite, defines a major structural and rheological transition in the lowermost mantle (171–173). The pressure and temperature conditions of this transition have been used to interpret seismic discontinuities (e.g., (174–176)), to calculate heat flow from the core (e.g., (177–179)), and to model the dynamics of deep mantle convection (e.g., (180,181)). The properties of the bridgmanite–post–perovskite transition varies among research groups in its thickness, pressure, Clapeyron slope $\left(\frac{dP}{dT}\right)$ (e.g., (174,175,182–184)), and EoSs (e.g., (22,185–192)).

The experimentally determined EoSs of pure $MgSiO_3$ end member bridgmanite and post–perovskite vary between studies, in part due to differences in pressure calibration materials (e.g., Au, MgO, Pt) and the pressure scale used. Figure 6a and 6b summarize the published EoSs for both bridgmanite and post–perovskite at 300 K. For example, (22,187,188,191) have reported EoS parameters calibrated using different MgO pressure scales (22,32,120,121). While (186,193,194) reported the EoS fit of bridgmanite based on the Pt pressure scales , (185,192) reported the EoS fit to bridgmanite using ruby scales (30,64), respectively. Ono et al., (190) and Ludin et al., (189) reports the EoS fit of post–perovskite and bridgmanite, respectively using Au scales (post–perovskite (74,81) and bridgmanite (85)). In terms of pressure, the variation between published bridgmanite and post–perovskite EoSs is ~10 GPa above 100 GPa. While other error sources are expected to contribute to this discrepancy, the 5–10% of the uncertainty partially arises from differences in pressure scales and must be accounted for.

[Figure 6 near here]

Using EoSAlign we align the EoSs of $MgSiO_3$ bridgmanite and post–perovskite; Figure 6. Figure 6e and 6f are the aligned data for bridgmanite and post–perovskite at 300 K, respectively. First, the pressure–volume conditions were copied from the literature, then the pressure values were entered in EoSAlign. The pressure scale used in the study was selected and aligned to Sakai et al., (22)'s MgO Keane EoS. When the pressure calibrant material was the same as the Sakai et al., (22), MgO, path 2a was used. For example, in the case of Komabayashi et al., (188), pressure was taken from their Table 1, then the pressure calibrant material ("composition" in the software) was selected as MgO, X–ray diffraction was selected for method, calibration was defined as Speziale et al., (120) ($K_0$=160.2 GPa), and then pressure was converted to Sakai et al., (22) (EoS= Keane).

All MgO EoSs were aligned to Sakai et al., (22) and the remaining studies using ruby, Pt and Au were aligned using path 2b. For example, in the case of Ono et al., (190) pressure was taken from their Figure/Table 2, the composition was selected as Au, method X–ray diffraction, calibration was defined as Anderson et al., (74), then "Select a study with a different composition and method to use for pressure calculation" was selected, the composition was MgO and method X–ray diffraction, then Sakai et al., (23) (Au) was selected followed by Sakai et al., (23) (MgO), finally pressure was converted to Sakai et al., (22) (EoS= Keane). The $MgSiO_3$ EoSs published with an MgO calibration are in good agreement with Sakai et al.,(22)'s and Tange et al., (191) $MgSiO_3$ EoSs. The $MgSiO_3$ EoSs published using Pt and Au vary slightly from the MgO calibrants when converted to the Sakai et al., (22) MgO scale. For post–perovskite Ono et al., (190) appears at higher pressure given the same volume; while for bridgmanite Komabayashi et al., (188) appears at higher pressure at the same volume.

[Table 3 near here]

We observed that the use of a common scale (Sakai et al., (22); Keane EoS, MgO) reduced the average discrepancy for bridgmanite by ~3 GPa at 80 GPa and by ~1 GPa for post–perovskite at 120 GPa. Using the aligned data we report updated fits for each polymorph using both 2$^{nd}$ and 3$^{rd}$ Birch–Murnaghan (BM2, BM3) and Keane formulations, based on the converted data to 160 GPa (Table 3). For bridgmanite and post–perovskite, we use all the data points shown in Figure 6; however, it should be noted that for post–perovskite, we capped the published data at 160 GPa to avoid over weighting the fit to one study. Sakai et al., (22) reports multiple EoSs for post–perovskite with data from their study and Guignot et al., (187) to ~260 GPa.

For bridgmanite, we fixed $K_0$ to 253 GPa from diffraction data in Vanpeteghem et al., (192) because the value was found to be in good agreement to Brillouin spectroscopy measurements (251−262 GPa) (195,196). $V_0$ of bridgmanite was fixed to 162.37 Å$^3$ because the value can be directly measured and has been reported for quenched samples at 1 bar to range from 162.0−162.75 Å$^3$ (197). The fits for bridgmanite, therefore, focus on constraining $K_0$'. Our $K_0$'s value is found to be 3.84 (2) for the 3$^{rd}$ order Birch Murnaghan fit. We found that the value for $K_0$' is heavily dependent on the results presented in two studies (188,193). When the data from Mao et al., (193) is omitted, then our $K_0$' is 4.09 (2) and when data from Komabayashi et al., (188) is omitted then our $K_0$' is 3.81 (2). This agrees well with low $K_0$' found in Mao et al., (193); where their $K_0$'=3.5, $K_0$ = 259.6 GPa, $V_0$ = 162.4 Å$^3$. The reason for the discrepancy between these two studies is not clear but we observe that this discrepancy shown in Figure 6g is not based on the pressure scale used.

For post–perovskite the uncertainty in published EoSs partially arises from it not being stable at 1 bar. Given pure post–perovskite transforms to glass on decompression, $V_0$ and $K_0$ are unknown. Our 3$^{rd}$ order Birch Murnaghan and Keane fits for post–perovskite were conducted for fixed $V_0$ value of 164.26 Å$^3$ from Sakai et al., (22) determined in their g-G plot. Since the uncertainty of the assumed $V_0$ value is difficult to assign, the uncertainty from our 3$^{rd}$ order Birch Murnaghan and Keane fits for post–perovskite is systematically underestimated. For post–perovskite, the pressure interval for the stability is very narrow for the Earth's mantle and the dataset we fit in this study (115−158 GPa). Given these limitations, we also fit a 2$^{nd}$ order Birch Murnaghan EoS with no free parameters. The advantage of this approach is to focus on the elastic properties for one parameter ($K_0$), which is the key for minimizing uncertainty and estimating physical properties (e.g. bulk sound speed) above 100 GPa.

By fitting EoSs to bridgmanite and post–perovskite we were able to constrain the isothermal bulk modulus and estimate the bulk sound speed ($V_\Phi$) of each phase. Because $V_\Phi$ is computed from the isothermal bulk modulus, this is a first–order estimate. However, by calculating a percent difference in $V_\Phi$ between two polymorphs we assume that the adiabatic correction $K_S = K_T(1 + \alpha\gamma T)$ for both phases is similar and cancels out. Using our 3$^{rd}$ and 2$^{nd}$ order Birch Murnaghan fit results for bridgmanite and post–perovskite, respectively, we found the difference in $V_\Phi$ across the bridgmanite–post–perovskite transition for $MgSiO_3$ is −0.68 ± 1.55% at 120 GPa. This decrease in $V_\Phi$ across the bridgmanite–post–perovskite transition agrees well with the observed difference at the D″ discontinuity below the Cocos Plate, −1.0 ± 0.5%, (198); Figure 7.

[Figure 7 near here]

However, our calculated decrease in $V_\Phi$ is less than the decrease observed in previous experimental measurements: Shieh et al., (199) obtained −2.3 ± 2.1% from their EoS of Fe bearing bridgmanite and post–perovskite (9 mol% Fe), and Yang et al., (200) measured a ~3.5% decrease (15 mol% Fe, 300 K) from compressional and shear wave velocities. This difference could be linked to the Fe within their samples or the density of the data set. We found that two studies (188,193) controlled the behavior of $K_0$'within our bridgmanite fit. These studies also impact the $V_\Phi$ across the transition. When we omit the recent dense dataset from Mao et al., (193), then our $V_\Phi$ is -3.1 ± 1.5 %. This results in a change in $V_\Phi$ across the transition that resembles previous experimental results (199,200). But when we omit data from Komabayashi et al., (188), then our $V_\Phi$ is -0.4 ± 1.6 % aligning better with (198) and other seismic studies that do not report a change in $V_\Phi$ at the D".

The aligned EoSs represents deviations between studies better than the unaligned EoSs. Yet, we found that the bridgmanite EoSs, $K_0$', and $V_\Phi$ at core–mantle boundary conditions are strongly influenced by a small number of data points that show substantial discrepancies. Because bridgmanite is quenchable, future studies should prioritize measurements above 100 GPa to resolve these discrepancies, reduce uncertainty in the change in $V_\Phi$ across the bridgmanite–post–perovskite transition, and better constrain the behavior of Earth-relevant compositions. More broadly, studies should continue to align data to a common pressure scale, particularly when comparing compositional effects on a mineral's EoS.

## 8. Conclusions

EoSAlign is an open–source software and a database designed for pressure calibration and the alignment of experimental pressure scales. The software is supported by an expanding database of 32 phases and over 400 calibrations spanning X–ray diffraction, Raman, and luminescence techniques. EoSHolo, a visualization module within EoSAlign, provides an interactive web for illustrating connections between pressure scales. In this work, we demonstrate that both the choice of pressure scale and its calibration history can lead to significant discrepancies between studies. We quantify pressure differences among six commonly used pressure scales (Au, $Al_2O_3$, diamond, Re, MgO, and Pt), with additional pressure scales provided in the supplementary material; we show that these discrepancies increase systematically with pressure. By using EoSAlign to align the bridgmanite and post–perovskite EoS, we present revised EoSs for these phases using Birch Murnaghan and Keane formulations. We observed that realigning datasets to a common pressure scale can reduce pressure–scale–related uncertainties for the EoSs of bridgmanite and post–perovskite in the $MgSiO_3$ system and better constrain material property changes at high pressures. However, when aligning data between different EoS "compositions" (materials) or methods, care must be taken to ensure that an appropriate conversion for pressure scale is used. When realigning Au pressure scales to a common Cu pressure scale (13), we found that the choice of pressure–conversion study affected the resulting alignment. Beyond the datasets examined here, EoSAlign can be used in the future to compare and align pressure scales across a range of experimental studies.

**Acknowledgements**

We thank the researchers who provided valuable feedback during discussions (S. Chen and E. O'Bannon) and during the Fall AGU 2025 meeting. This input helped improve the software and expand the compositional range examined in this study.

***Funding Details***

This work was supported by the NASA, grant (80NSSC25K7021).

***Disclosure Statement***

The authors report there are no competing interests to declare.

***Declaration of generative AI use***

We acknowledged generative AI tools were used during the preparation of this work. ChatGPT (OpenAI) was used to assist with literature metadata extraction, Claude (Anthropic) was used to assist with software debugging and design, and both were used to aid in spelling and grammar. Each tool was used throughout the code and manuscript development between Augusts 2025 and July 2026, over which multiple successive versions of each tool were released and used. We reviewed, checked, and verified all AI-generated suggestions, and we confirm the originality and accuracy of the content.

***Data and Software Availability***

The EoSAlign software used for this publication can be accessed on Zenodo (201) and at GitHub via the EoSAlign repository within the EoSApplications organization (https://github.com/EoSApplications).

***Supplemental Online Material***

The online supplemental material can be found on Zenodo (202), in the supplement we provide a summary document of each pressure scale exported and the corresponding reference scale. In separate folders, we provide figures for each pressure scale and a corresponding .csv file for each combination detailed in the table. All figures can be viewed online or generated by the user in EoSAlign.

**Tables**

Table 1: Summary of the equations of state calibrations available in EoSAlign.

| Composition | Studies |
|---|---|
| Ag | (18,28,34,51–56) |
| Al | (12,16,19,29,34,52,56–59) |
| Al2O3 | (1,14,18,19,25,26,28–36,53,56,58,60–70) |
| Ar | (5,16,71,72) |
| Au | (2,3,15,16,18,19,23,29,34,54,56,58,73–88) |
| Be | (42,89–92) |
| Bi | (73,93,94) |
| Co–hcp | (28,95–98) |
| Cu | (12–15,18,19,23,29,34,52–54,56,58,92,99) |
| Diamond | (17,19,23,33,34,38–41,56,100,101) |
| Fe–bcc | (28,82) |
| Fe–hcp | (23,63,82,102–108) |
| He | (109,110) |
| KCl–B2 | (111–115) |
| MgO | (22,23,32,34,52,56,64,76,77,81,82,86,105,116–121) |
| Mo | (15,23,28,52,56,75,82,88,122–125) |
| NaCl–B1 | (76,82,126–130) |
| NaCl–B2 | (1,2,23,75,76,82,128,131–135) |
| Nb | (56) |
| Ne | (1,2,5,17,75,136) |
| Ni | (28,104,137,138) |
| Pd | (52,139–142) |
| Pt | (2,10,15,19,23,29,34,56,58,75–78,81,82,86,87,143,144) |
| Re | (6,23,43,44,145–151) |
| SrB4O7 | (37,70,152–155) |
| Ta | (12,15,19,34,56,58,82,156–162) |
| W | (12,19,23,29,34,56,58,82,123,163–165) |
| YAG | (166) |
| YAG–Y1 | (70,166) |
| YAG–Y2 | (70,166) |
| Zn | (28,167–169) |
| cBN | (45–50,170) |

Table 2: These equations of state are listed as category 3, the published study reported the reduced isotherm or raw data without an isothermal equation of state. We fit and report the Birch–Murnighan EoS fit to the published data that will appear unrounded in EoSAlign. It is not recommended to compare studies with this study above the maximum pressure as literature may have fit this study differently. *Sodium and Mg are not compositions currently supported by EoSAlign, they are included for completeness (52). Error assigned to shock studies represents the uncertainty for the EoS from the reduced shock isotherm, it does not account for total experimental uncertainty.

| Study | Composition | P0 (GPa) | $V_0$ (cc/mol) | $K_0$ (GPa) | $K_0'$ | Max Pressure (GPa) |
|---|---|---|---|---|---|---|
| (52) | Al | 0 | 9.6882 | 77.11 (6) | 4.19 (3) | 110 |
| (52) | Cu | 0 | 7.116 | 132.2 (3) | 5.19 (1) | 160 |
| (52) | Mo | 0 | 9.4013 | 267.57 (4) | 3.8153 (9) | 200 |
| (52) | Ag | 0 | 10.283 | 101.1 (5) | 5.85 (3) | 150 |
| (52) | Pd | 0 | 8.875 | 177.4 (5) | 5.7 (2) | 200 |
| (52) | Na* | 0 | 23.749 | 5.59 (5) | 4.43 (3) | 35 |
| (52) | Mg* | 0 | 13.968 | 33.60 (9) | 4.134 (9) | 70 |
| (52) | MgO | 0 | 11.2424 | 151.8 (1) | 4.59 (6) | 120 |
| (81) | Au | 0 | 10.2124 | 177.3 (6) | 4.77 (3) | 83 |
| (81) | Pt | 0 | 9.0904 | 270.72 (7) | 5.267 (3) | 91 |
| (81) | MgO | 0 | 11.2424 | 154.02 (9) | 4.544 (4) | 93 |
| (72) | Ar | 0 | 21.343 ± 3.78 | 5.70 (6.0) | 5.5 (1.2) | 80.6 |
| (110) | He | 15.6 | 3.944 | 34.1 (1.6) | 6.7 (7) | 23.3 |
| (127) | NaCl–B1 | 0 | 27.012 | 25.03 | 4.521 (5) | 31 |
| (123) | W | 0 | 9.5466 | 311.2 (1) | 3.895 (2) | 300 |
| (123) | Mo | 0 | 9.393049437 | 266.8 (2) | 3.867 (3) | 300 |

Table 3: EoS of bridgmanite and post–perovskite refit to aligned data on the $MgSiO_3$ system. Fixed values are denoted with *. For Bridgmanite $V_0$ was fixed to 162.37 from (191) and $K_0$ was fixed to 253 GPa from (192). For post–perovskite $V_0$ was fixed to 164.26 from (22) and $K'_\infty$ was fixed to 2.19 from (22).

| Polymorph | Scale | Equation | $V_0$ | $K_0$ | $K'_0$ | $K'_\infty$ | $P_0$ |
|---|---|---|---|---|---|---|---|
| Bridgmanite | MgO (22) Keane EoS | BM3 | **162.37*** | **253*** | 3.84 (2) | | 0 |
| | | Keane | **162.37*** | **253*** | 4.06 (3) | **2.19*** | 0 |
| Post–perovskite | | BM3 | **164.26** | 217 (10) | 4.2 (3) | | 0 |
| | | Keane | **164.26*** | 197 (16) | 5.7 (9) | **2.19*** | 0 |
| | | BM2 | 123.1 (1) | 636 (20) | | | 115 |

## Figures

Figure 1:

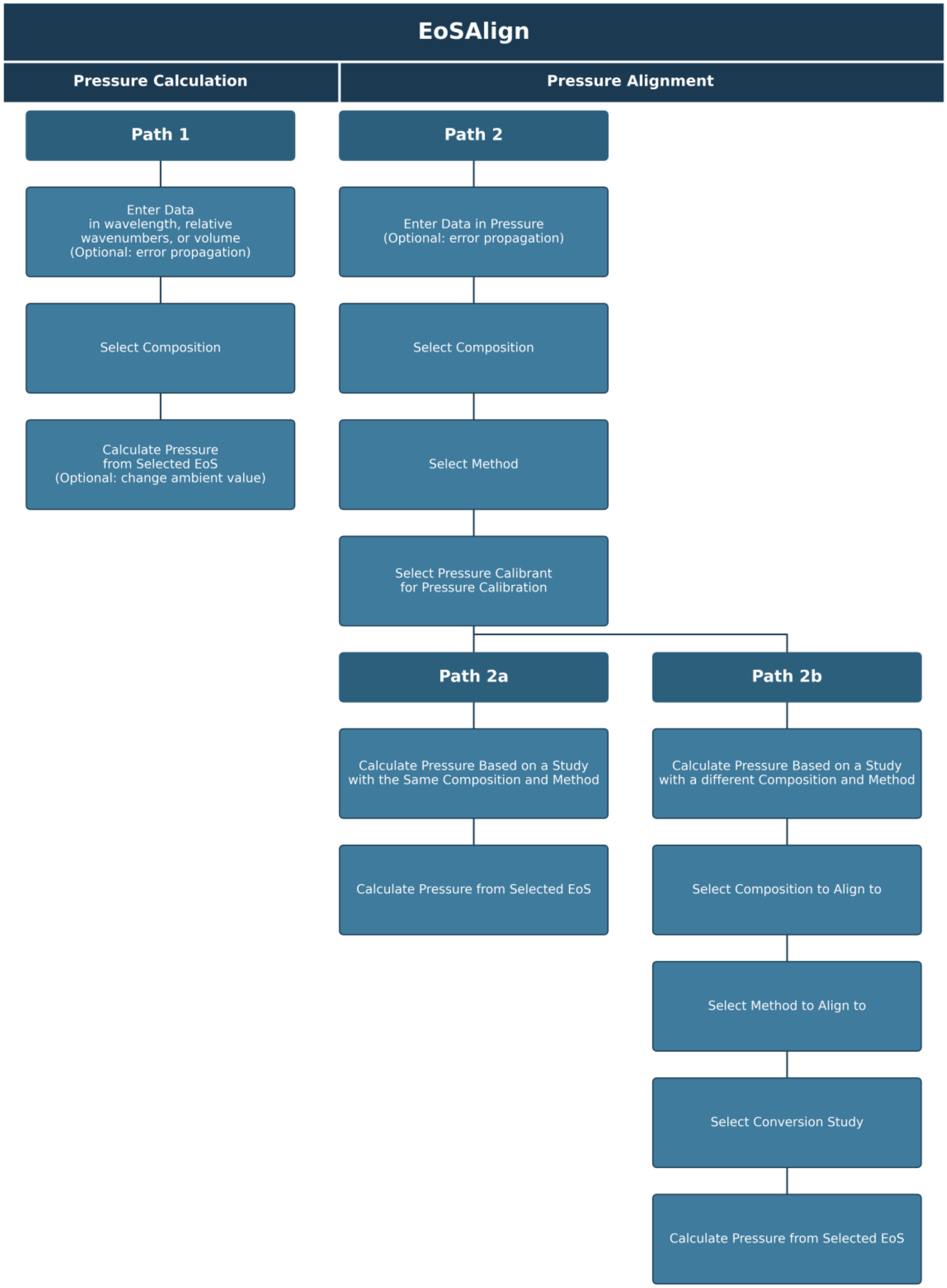

Figure 2:

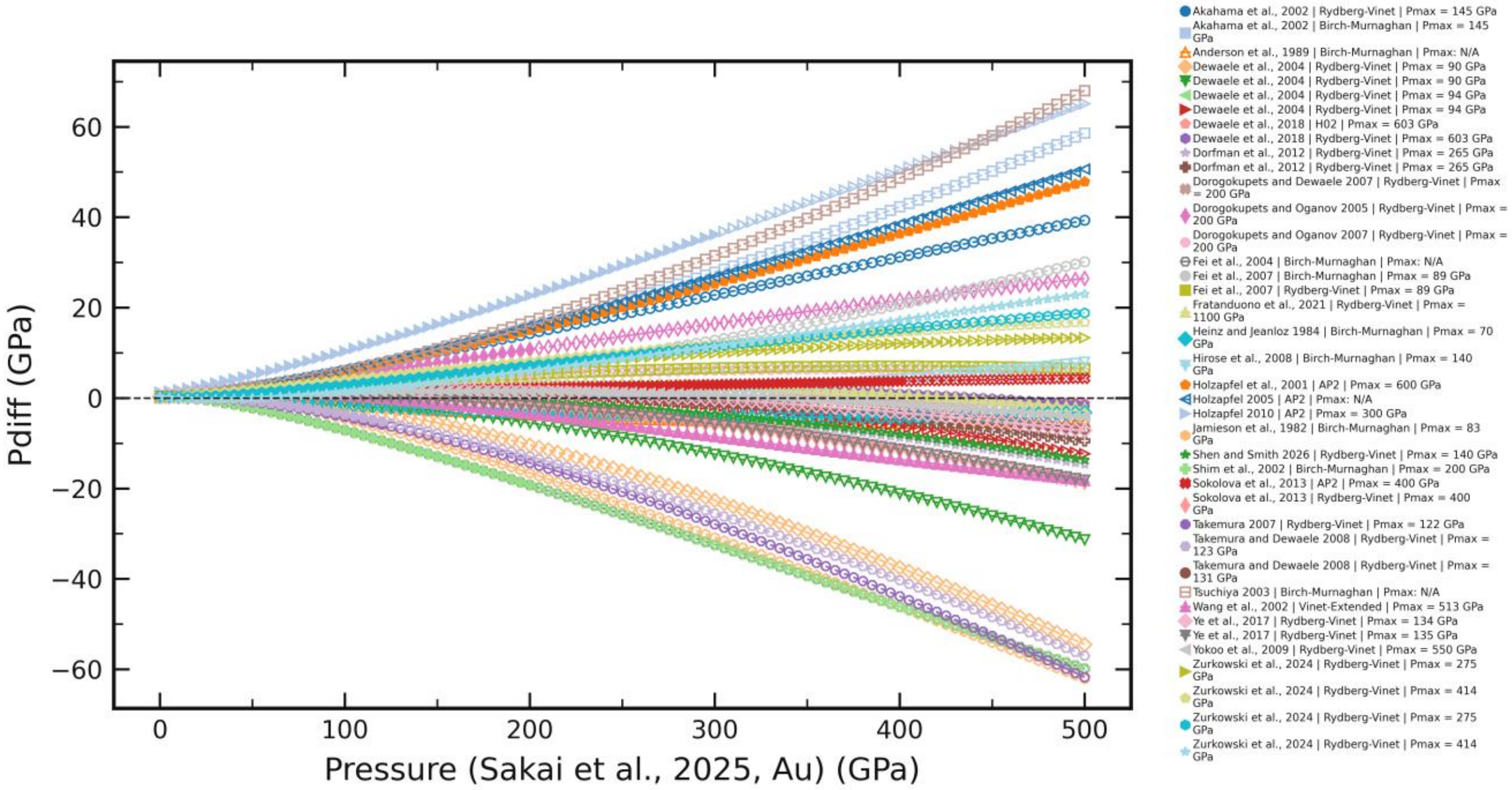

Figure 3:

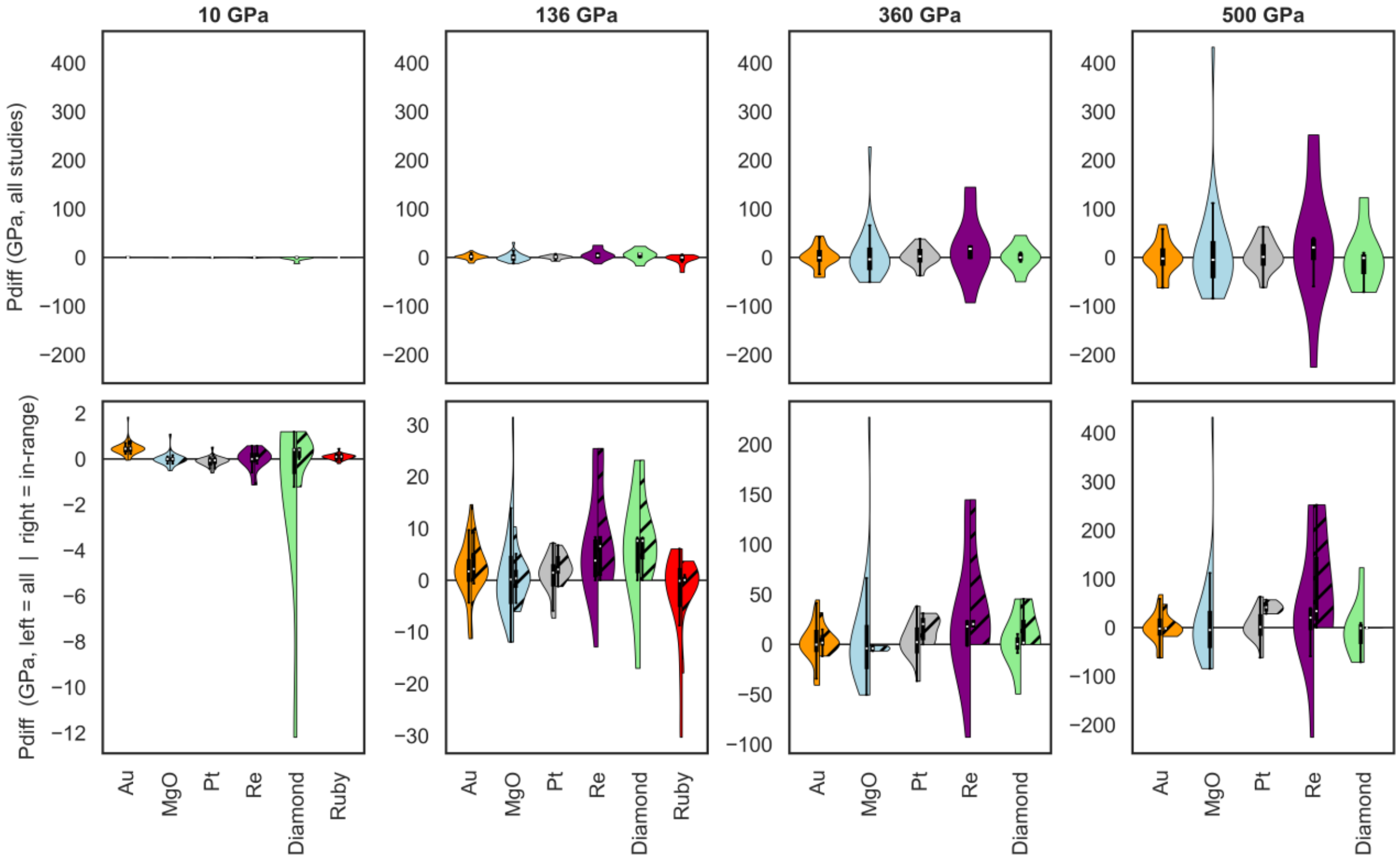

Figure 4:

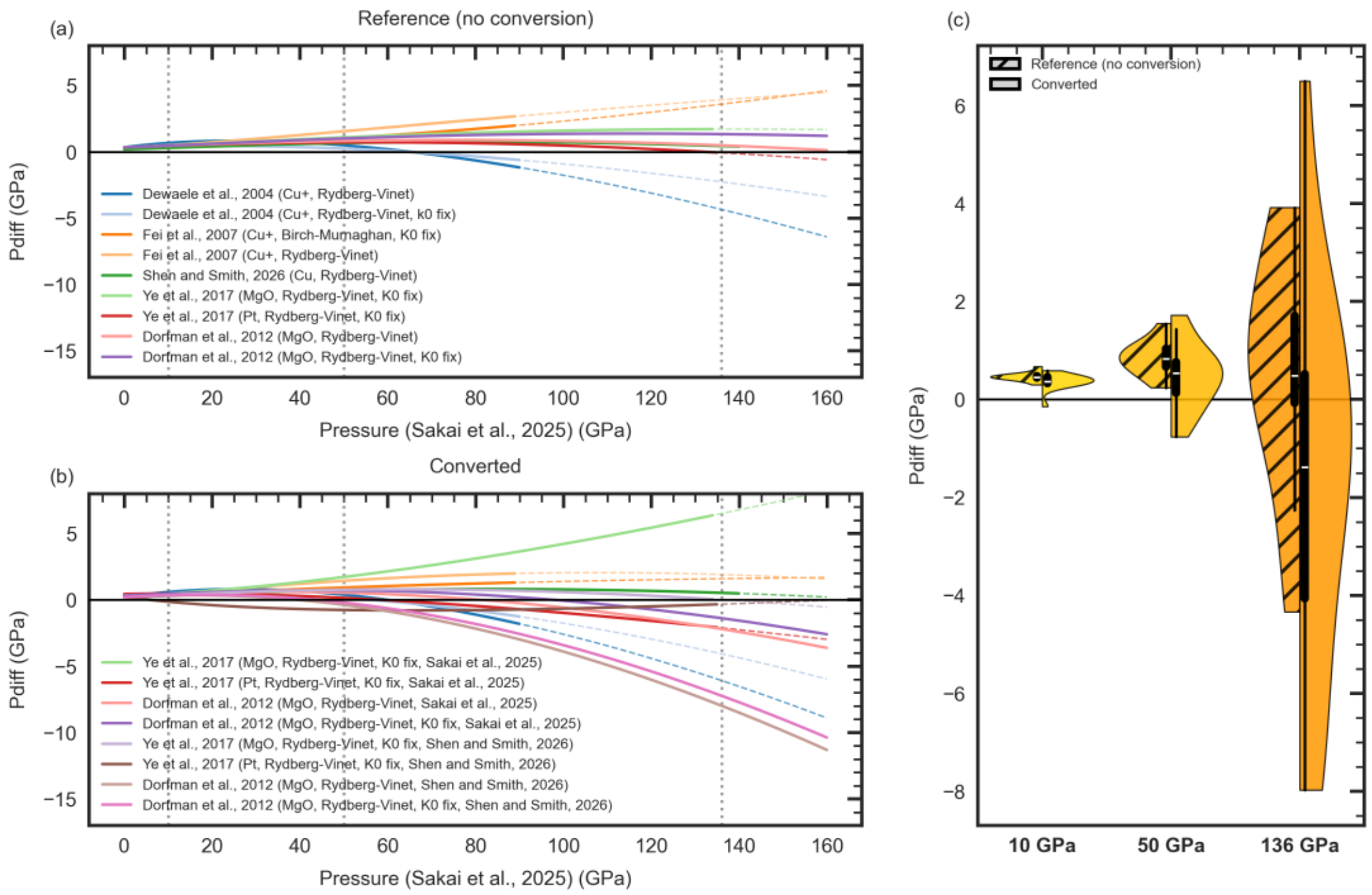

Figure 5:

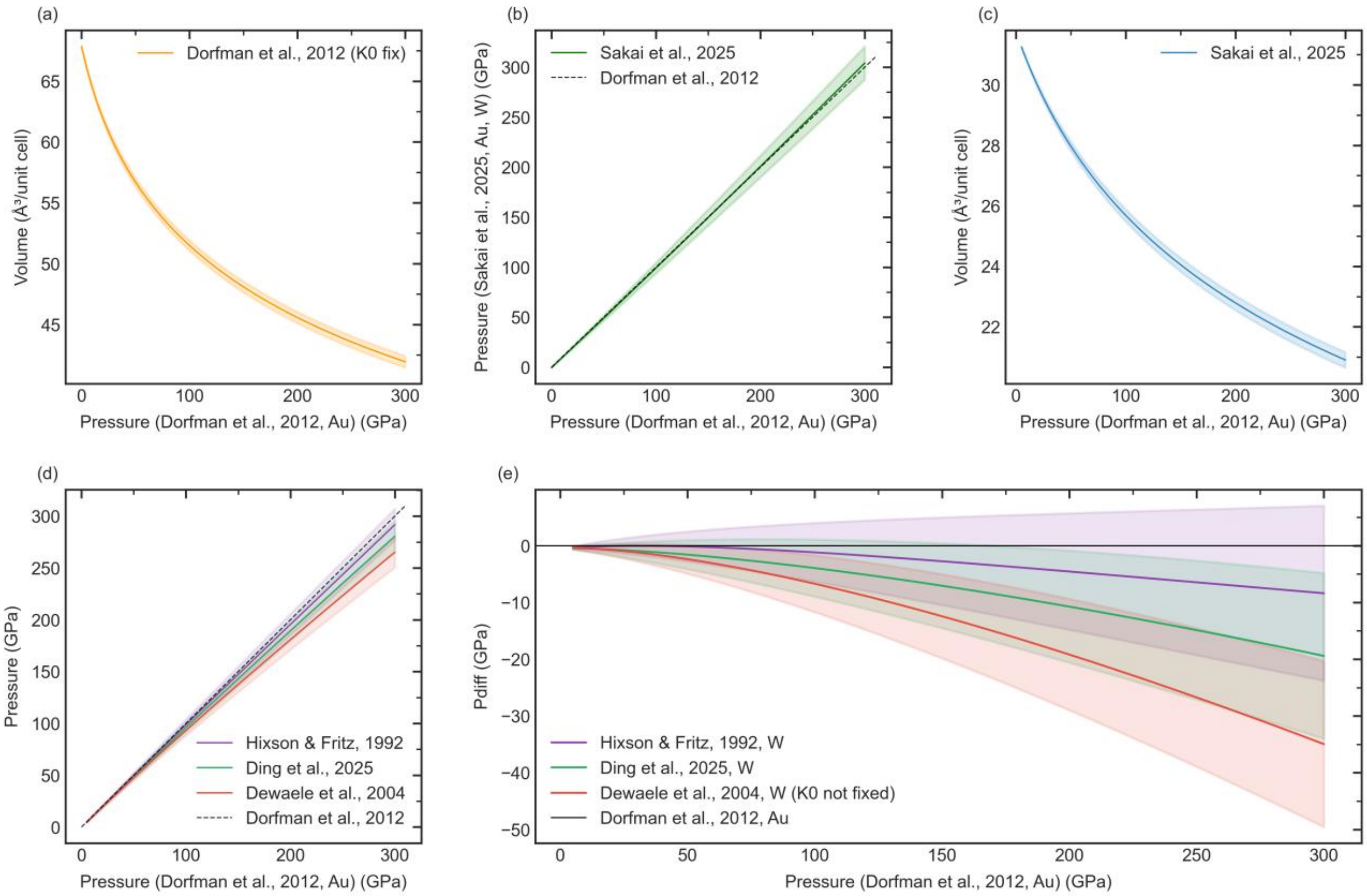

Figure 6:

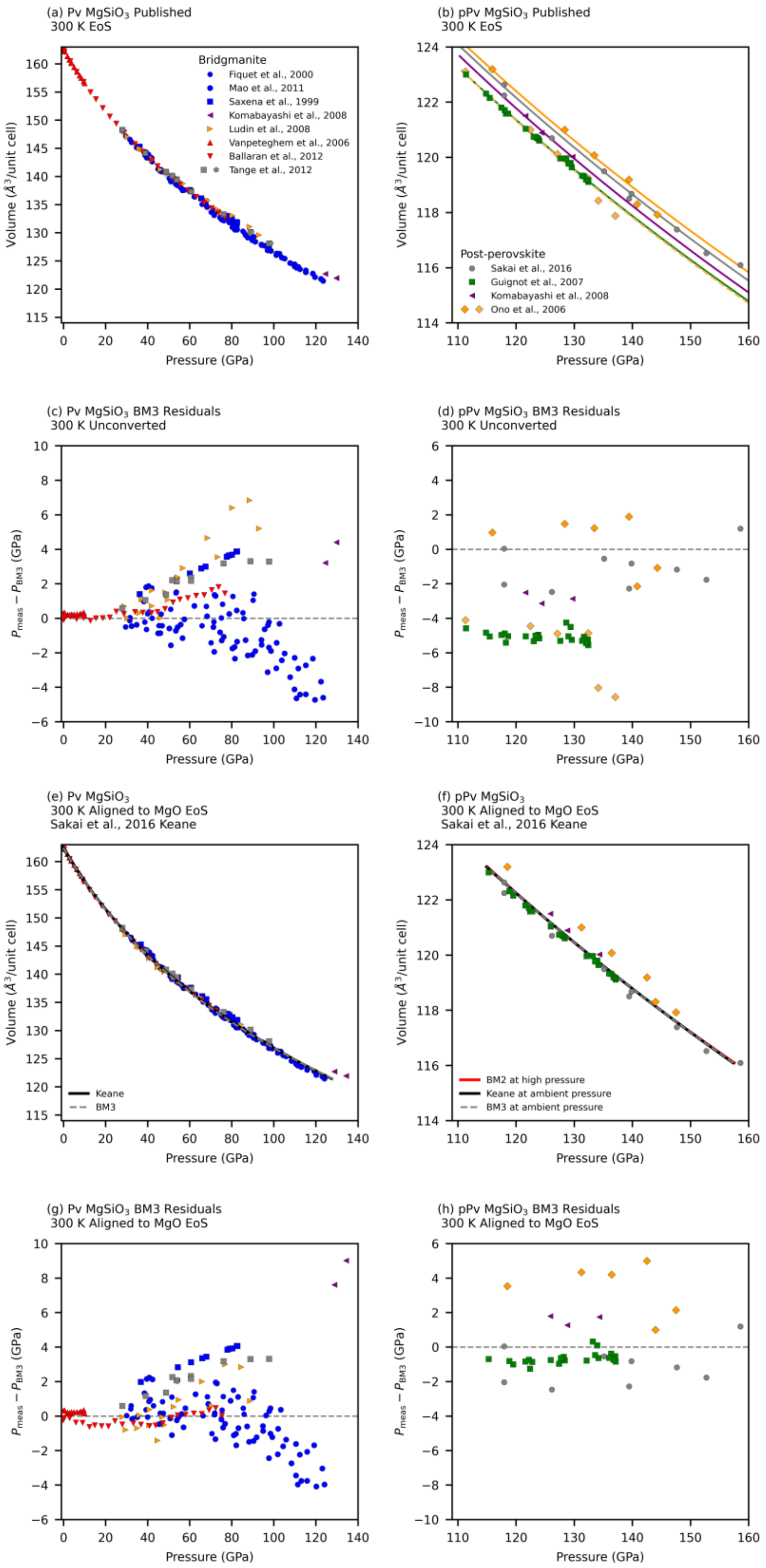
(a) Pv MgSiO$_3$ Published
300 K EoS
Bridgmanite
Fiquet et al., 2000
Mao et al., 2011
Saxena et al., 1999
Komabayashi et al., 2008
Ludin et al., 2008
Vanpeteghem et al., 2006
Ballaran et al., 2012
Tange et al., 2012
Volume ($Å^3$/unit cell)
Pressure (GPa)
(b) pPv MgSiO$_3$ Published
300 K EoS
Post-perovskite
Sakai et al., 2016
Guignot et al., 2007
Komabayashi et al., 2008
Ono et al., 2006
Volume ($Å^3$/unit cell)
Pressure (GPa)
(c) Pv MgSiO$_3$ BM3 Residuals
300 K Unconverted
$P_{meas} - P_{BM3}$ (GPa)
Pressure (GPa)
(d) pPv MgSiO$_3$ BM3 Residuals
300 K Unconverted
$P_{meas} - P_{BM3}$ (GPa)
Pressure (GPa)
(e) Pv MgSiO$_3$
300 K Aligned to MgO EoS
Sakai et al., 2016 Keane
Keane
BM3
Volume ($Å^3$/unit cell)
Pressure (GPa)
(f) pPv MgSiO$_3$
300 K Aligned to MgO EoS
Sakai et al., 2016 Keane
BM2 at high pressure
Keane at ambient pressure
BM3 at ambient pressure
Volume ($Å^3$/unit cell)
Pressure (GPa)
(g) Pv MgSiO$_3$ BM3 Residuals
300 K Aligned to MgO EoS
$P_{meas} - P_{BM3}$ (GPa)
Pressure (GPa)
(h) pPv MgSiO$_3$ BM3 Residuals
300 K Aligned to MgO EoS
$P_{meas} - P_{BM3}$ (GPa)
Pressure (GPa)

Figure 7:

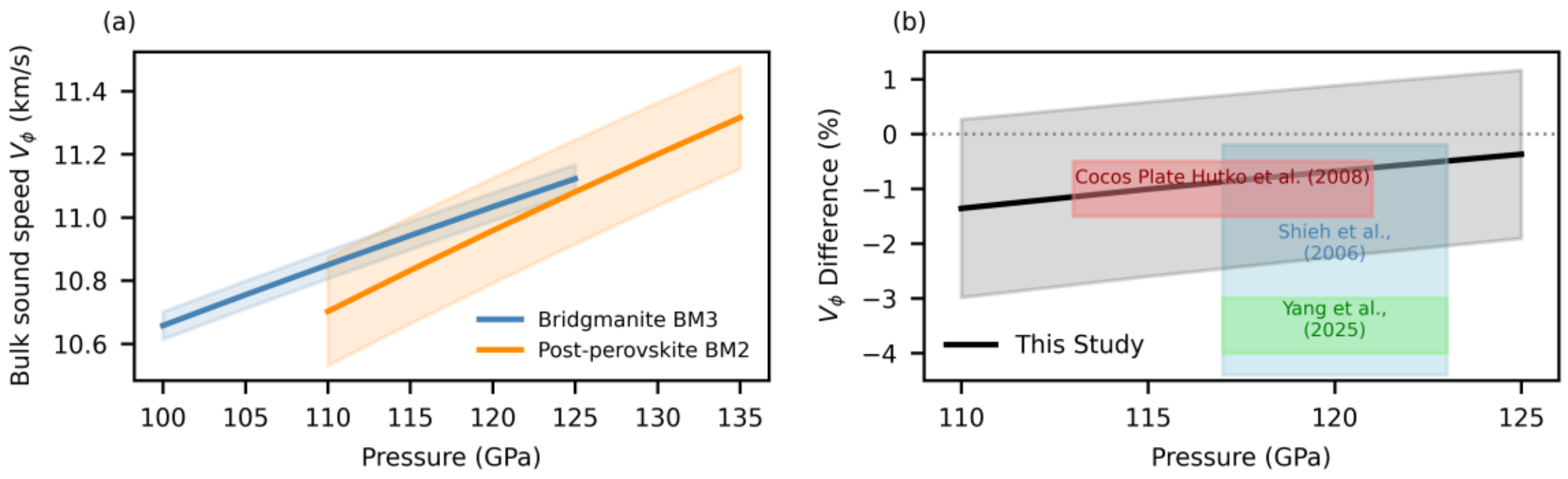

## Figure Captions

**Figure 1:** EoSAlign has three standard paths depending on if the objective is pressure calculation or pressure alignment/comparison. The first path enables a user to calculate pressure from a measured volume, Raman shift, or luminescence peak position. The second path is divided into two paths, path 2a enables the user to compare and align pressure from literature studies of the same composition, while path 2b enables a user to compare pressure from literature studies with different compositions.

**Figure 2:** Example of a pressure difference ($P_{diff}$) figure exported from EoSAlign for all Au pressure scales. This figure was calculated using path 2a of EoSAlign where the input pressure was in steps of 5 GPa from ambient to 500 GPa. The input pressure was calibrated to Sakai et al., (23) using the Rydberg–Stacey EoS formulation. Filled symbols are within the pressure range of the datasets and open symbols are for extrapolations. Studies that do not have a defined dataset coverage for pressure are open symbols.

**Figure 3:** Violin plots of commonly used pressure scales at 10, 136, 360, and 500 GPa. The y–axis is the pressure difference between a given study–the reference study. The reference study used here is Sakai et el., (23) for all compositions except Ruby where Shen et al., (35) was used. The top row has each of the violin plots scaled to a common y–axis range while the bottom row has each violin plot scaled to different y–axis scales. The bar within the violin plot provides statistical information on the range of each pressure scale: the white dot is the median, the thick black bar is the interquartile range (25$^{th}$ to 75$^{th}$ percentile, IQR), while the thin black line is 1.5 IQR. The solid part of the violin is calculated using all studies in EoSAlign. The hatched part of the violin plot represents only unextrapolated pressure scales at a given pressure range. An exception is for the 500 GPa column, the hatched region includes studies with a maximum pressure greater than 430 GPa no violin plot appears when less than 2 studies in EoSAlign have a maximum pressure greater than 430 GPa.

**Figure 4**: Comparison of intercalibrated Au pressure scales. a) pressure difference (pdiff Study–Sakai et al., (23)) for each Au pressure scale as published. b) pressure difference for each Au pressure scale realigned to a common Cu reference scale (13). c) violin plot of the Au pressure scale with no conversion with hatched lines and converted realigned Au pressure scales. The bar

within the violin plot provides statistical information on the range of each pressure scale, the white region is the median, the thick black bar is the interquartile range ($25^{th}$ to $75^{th}$ percentile, IQR), while the thin black line is 1.5 IQR.

**Figure 5:** Visualization of error bar expansion for path 2b converting between an input pressure for Au from Dorfman et al., (75), Rydberg–Vinet EoS, with $K_0$ not fixed through Sakai et al., (23) to convert from Au to W to three different W calibrants ((29) with $K_0$ not fixed, (123,163)). a). calculated volume and uncertainty attributed to Au after converting from Dorfman et al., (75) assuming a starting 5% pressure uncertainty. b). Calculated pressure and uncertainty for Au and W based on the Sakai et al., (23) pressure scale, EoSAlign assumes this is equal to the pressure and uncertainty for their W pressure scale. c). calculated volume and uncertainty for W based on the Sakai et al., (23) W pressure scale. d). Calculated pressure and uncertainty for W based on three select pressure scales. e). Pressure difference for the three – Dorfman et al., (75) (Au) vs Dorfman et al., (75) Au pressure. These W scales result in lower pressure than Au and the pressure uncertainty at 250 GPa (12.5 GPa) increased by ~0.5–1 GPa depending on the W calibrant used.

**Figure 6:** EoSs of $MgSiO_3$ bridgmanite and post–perovskite to 160 GPa. a) EoSs of bridgmanite with data in the literature. b) EoSs of post–perovskite with data and EoSs published in the literature. e) The EoSs of bridgmanite aligned to the MgO EoS proposed in Sakai et al., (22) using data from Tange et al., (121) and a Keane fit to the data. f). The EoS of post–perovskite aligned to the MgO EoS proposed in Sakai et al., (22). This study fits an EoS to bridgmanite and post–perovskite using the Keane EoS (solid black) and a $2^{nd}$ and $3^{rd}$ order Birch Murnaghan EoS (red BM2, dashed grey BM3). c,d,g,h) residuals of the unconverted or aligned data when compared to the $3^{rd}$ order Birch Murnaghan fits to the data for bridgmanite and post–perovskite.

**Figure 7:** Bulk sound speed of the rederived EoSs for bridgmanite and post–perovskite at 300 K. a). Bulk sound speed of bridgmanite (blue) and post–perovskite (orange) to 160 GPa. The shaded area are propagated uncertainties from our BM3 and BM2 fits for the phases, respectively. b) the difference in bulk sound speed. The range in the seismologically measured bulk sound speed for the D" discontinuity at the Cocos plate is shown in red (198) while experimental measurements are shown in blue (199) and green (200) for Fe bearing bridgmanite and post–perovskite.